\documentclass[11pt]{article}
\usepackage[preprint]{acl}
\usepackage{times}
\usepackage{latexsym}
\usepackage[T1]{fontenc}
\usepackage[utf8]{inputenc}
\usepackage{microtype}
\usepackage{graphicx}
\usepackage{float}
\usepackage{xcolor}
\usepackage{colortbl}
\usepackage{booktabs}
\usepackage{array}
\usepackage{tabularx}
\usepackage{multirow}
\usepackage{amsmath}

\definecolor{DataColor}{HTML}{AD245E}
\definecolor{CiteColor}{HTML}{2563A6}
\definecolor{FindingColor}{HTML}{A55314}
\definecolor{FigureColor}{HTML}{527D9F}
\definecolor{FigureFill}{HTML}{F2F6FA}

\title{Jev in the Wild: A Data-Driven Analysis of the Jev Model's\\ Functionality, Applications and Ecosystem}
\author{Guoming Ling\textsuperscript{1},
  Muen Xue\textsuperscript{1},
  Zijian Ye\textsuperscript{2} \\
  \textsuperscript{1}Sun Yat-sen University,
  \textsuperscript{2}The Chinese University of Hong Kong}

\begin{document}
\maketitle

\begin{abstract}
Jev is a fast, low-cost decision model that answers natural-language questions with choices, binary judgments, and scores. As its public ecosystem grows rapidly, it remains unclear how Jev is used across applications and how public attention relates to project distribution. To answer these questions, we conduct a large-scale, data-driven analysis of 2,170 publicly available Jev projects collected from GitHub as of September 22, 2026. We find rapid early growth in Jev's public ecosystem, with both new projects and integration into existing repositories. Across diverse domains, projects use Jev for multiple decision purposes and combine its interfaces. Attribute judgment and scoring are widely used, while the use of action selection, content filtering, and model and tool selection varies across domains. These patterns suggest that Jev serves as a reusable decision component whose functionality varies with the surrounding workflow. Meanwhile, public attention is concentrated in routing and interface agents and does not track project counts. Our findings provide a quantitative view of Jev's emerging ecosystem and inform the design and evaluation of general-purpose decision models across diverse application contexts.
\end{abstract}

\section{Introduction}
\label{sec:intro}

Jev is a recently introduced model for fast, low-cost decisions defined through natural language \cite{typesafe_api}. Developers specify a decision question and its possible outputs, and Jev returns a choice, probability, or score. It provides three interfaces: \texttt{Choice} selects among candidate options, \texttt{Noul} makes binary judgments, and \texttt{Score} rates an input against predefined levels \cite{almeida2026jev}. Together, these interfaces support lightweight decisions such as model routing, context filtering, and action selection \cite{li2026replacinglargelanguagemodels, jiang2026jevmemsystemonecontrolledagenticmemory}.

\paragraph{Why is Jev useful?}
Many applications require frequent, lightweight decisions \cite{mozzarelli2026ai}, but existing models do not fully meet this need. Traditional supervised classifiers require task-specific labeled data and training, and changes to the label space often require collecting new data and retraining the model \cite{devlin-etal-2019-bert}. LLMs provide much stronger zero-shot generalization and avoid task-specific training \cite{NEURIPS2020_1457c0d6}, but their autoregressive generation introduces substantial latency and inference cost \cite{NEURIPS2023_7b97adea}, especially for simple decisions with a small output space. Jev targets this gap by combining zero-shot generalization with fast and low-cost inference \cite{ibrahim2026evaluatingdecisionmodelstext}. This makes it suitable for applications that require many small, diverse, and frequently changing decisions.

\begin{figure}[t]
  \centering
  \includegraphics[width=\columnwidth]{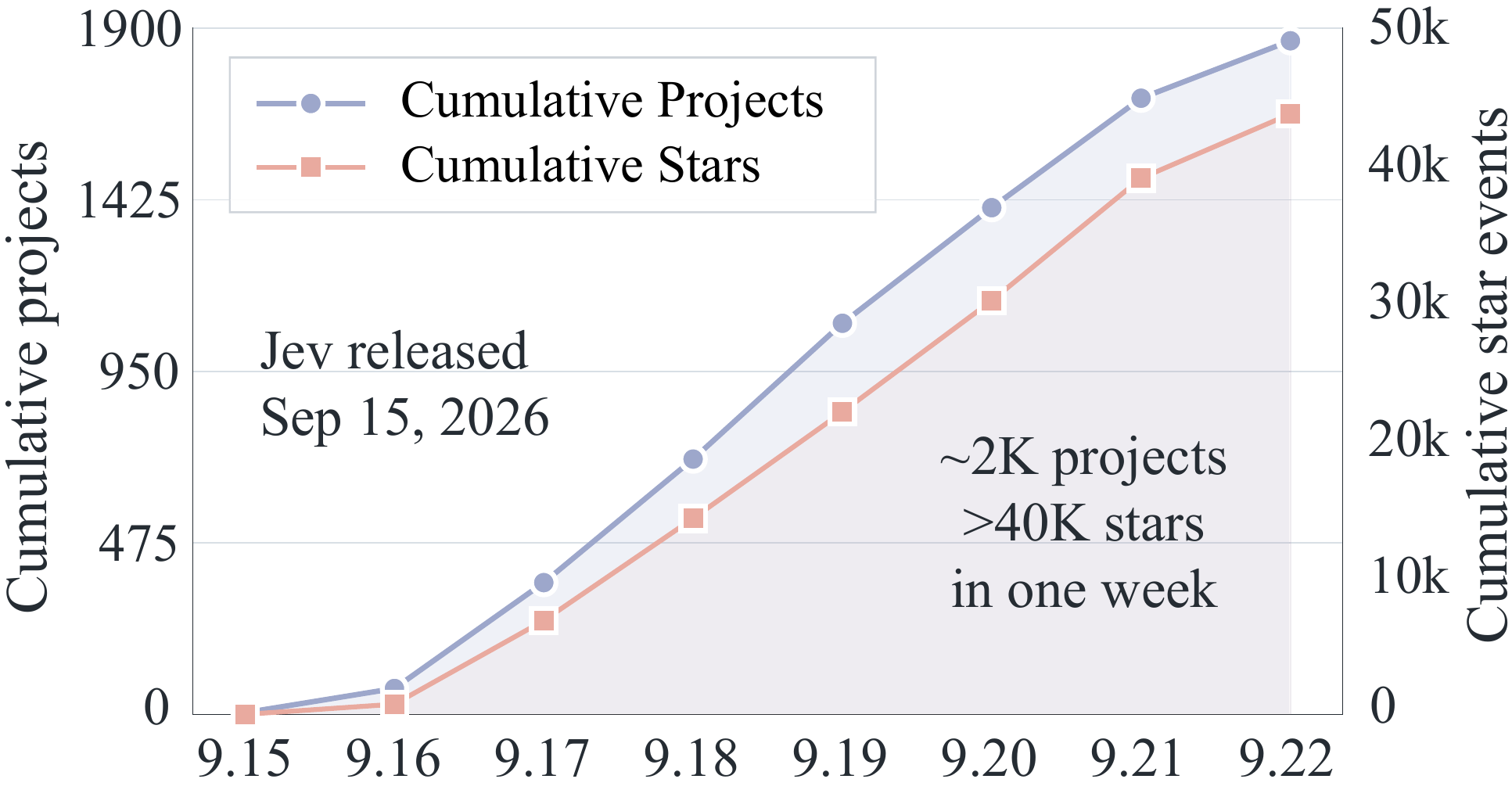}
  \vspace{-18pt}
  \caption{\textbf{Jev's ecosystem grew explosively} in its first week after release on September 15, 2026, expanding to 1,865 new GitHub repositories and gaining 43,750 stars. This rapid rise demonstrates strong early momentum.}
  \label{fig:growth}
\end{figure}

\paragraph{How is Jev used in practice?}
Jev is typically used as a lightweight decision component within a larger system \cite{jiang2026jevmemsystemonecontrolledagenticmemory}. The surrounding application prepares the relevant state and defines the decision, while Jev evaluates the request and returns the result \cite{li2026replacinglargelanguagemodels}. The system then uses this result to control the next operation \cite{wu2026reflexjevefficientselective}. In this way, Jev handles the decision itself, while the surrounding system manages task-specific context and execution.

\begin{table*}[t]
\centering
\normalsize
\setlength{\tabcolsep}{5pt}
\renewcommand{\arraystretch}{1.04}
\begin{tabularx}{\textwidth}{@{}ll*{4}{>{\raggedleft\arraybackslash}X}@{}}
\toprule
\rowcolor{FigureFill}\textbf{Major category} & \textbf{Subcategory} & \textbf{Projects} & \textbf{\% of total} & \textbf{Stars} & \textbf{\% of stars} \\
\midrule
Interface Agents & Web Interaction & 125 & 5.8\% & 46,067 & 21.0\% \\
 & Desktop \& Mobile & 50 & 2.3\% & 1,418 & 0.6\% \\
\midrule
Software Engineering & Code Quality & 90 & 4.1\% & 1,500 & 0.7\% \\
 & Dev Workflows & 218 & 10.0\% & 3,014 & 1.4\% \\
\midrule
Search \& Memory & Context \& Memory & 77 & 3.5\% & 10,533 & 4.8\% \\
 & Search \& Data & 120 & 5.5\% & 2,334 & 1.1\% \\
 & Classification & 184 & 8.5\% & 1,169 & 0.5\% \\
\midrule
Safety \& Governance & Review \& Approval & 82 & 3.8\% & 5,750 & 2.6\% \\
 & Security \& Compliance & 93 & 4.3\% & 1,904 & 0.9\% \\
\midrule
Routing \& Automation & Model \& Tool Routing & 132 & 6.1\% & 71,671 & 32.6\% \\
 & Workflow Automation & 118 & 5.4\% & 19,332 & 8.8\% \\
\midrule
Simulation \& Control & Games \& Simulation & 227 & 10.5\% & 1,825 & 0.8\% \\
 & Robotics \& Control & 25 & 1.2\% & 264 & 0.1\% \\
\midrule
Content \& Expert Tasks & Professional Tasks & 176 & 8.1\% & 6,597 & 3.0\% \\
 & Content \& Dialogue & 211 & 9.7\% & 5,363 & 2.4\% \\
\midrule
Infrastructure \& Other & SDKs \& Tools & 175 & 8.1\% & 34,800 & 15.8\% \\
 & Research \& Resources & 56 & 2.6\% & 6,010 & 2.7\% \\
 & Other Applications & 11 & 0.5\% & 107 & $<0.1\%$ \\
\bottomrule
\end{tabularx}
\caption{Distribution of 2,170 public Jev projects across 8 major categories and 18 subcategories. Public Jev projects span a broad range of application domains, showing that \textbf{Jev is already being used for diverse decision tasks}.}
\label{tab:domains}
\end{table*}

As shown in Figure~\ref{fig:growth}, Jev has rapidly grown from a new model into an emerging ecosystem of public projects. However, it remains unclear how this ecosystem is evolving and how Jev is actually used across applications. To answer these questions, we conduct a large-scale, data-driven analysis of 2,170 publicly available Jev projects collected from GitHub as of September 22, 2026. We systematically examine the growth of the ecosystem, the applications and decision roles of Jev, and how public attention is distributed across these applications. Our analysis provides a quantitative view of the emerging Jev ecosystem and a reference for understanding how the model is being adopted in practice. We discuss related work in Appendix \ref{app:related_work} and summarize our contribution:
{\setlength{\leftmargini}{1.0em}
\begin{itemize}
    \item \textbf{Growth trends.} We quantify the growth of public Jev projects and GitHub attention over time, revealing rapid and bursty ecosystem growth.
    \item \textbf{Usage patterns.} We characterize how Jev is used through application domains, decision purposes, and interface adoption, and compare project distribution with public attention.
\end{itemize}}

\section{Data Collection and Growth Trends}
\label{sec:data}

In this section, we first describe the dataset and then analyze project growth and community attention.

\subsection{Data Collection}
\label{sec:collection}

We construct the dataset in three stages: candidate retrieval, project verification, and annotation.

\paragraph{Candidate retrieval.}
We collect candidate repositories from GitHub using predefined search queries based on Jev-related keywords, API endpoints, and SDK references. We combine repository and code search and deduplicate the retrieved repositories using GitHub repository IDs.

\paragraph{Project verification.}
Each candidate repository is inspected by a GPT-6 Luna Max agent. We include a repository only when its public code or documentation provides clear evidence that Jev is used for a concrete task. Repositories that only mention Jev or provide generic wrappers or SDKs without a concrete application are excluded.

\paragraph{Annotation.}
For each included project, a GPT-6 Luna Max agent annotates its primary application domain, decision purposes, interface usage, and other attributes required by our analyses. Application domain describes what the project is built for, while decision purpose describes the role performed by Jev within the project. Key inclusion decisions and application-domain labels are independently reviewed by a second GPT-6 Luna Max agent, and disagreements are resolved by inspecting the original repository materials. This process yields 2,170 verified Jev projects for our analysis.

\subsection{Growth Trends}
\label{sec:growth}

Figure~\ref{fig:growth} shows that Jev was adopted rapidly after its release on September 15, 2026. Of the 2,170 repositories in our dataset, 1,865 were created in the following week, while 305 existing repositories integrated Jev. This pattern shows that Jev attracted both new development and adoption by established projects.
Public attention increased just as quickly. During the same week, these new repositories gained 43,750 stars, rapidly drawing public attention. Growth in both repositories and stars continued beyond this initial surge, indicating strong and sustained momentum across the Jev ecosystem.

\begin{figure}[t]
  \centering
  \includegraphics[width=\columnwidth]{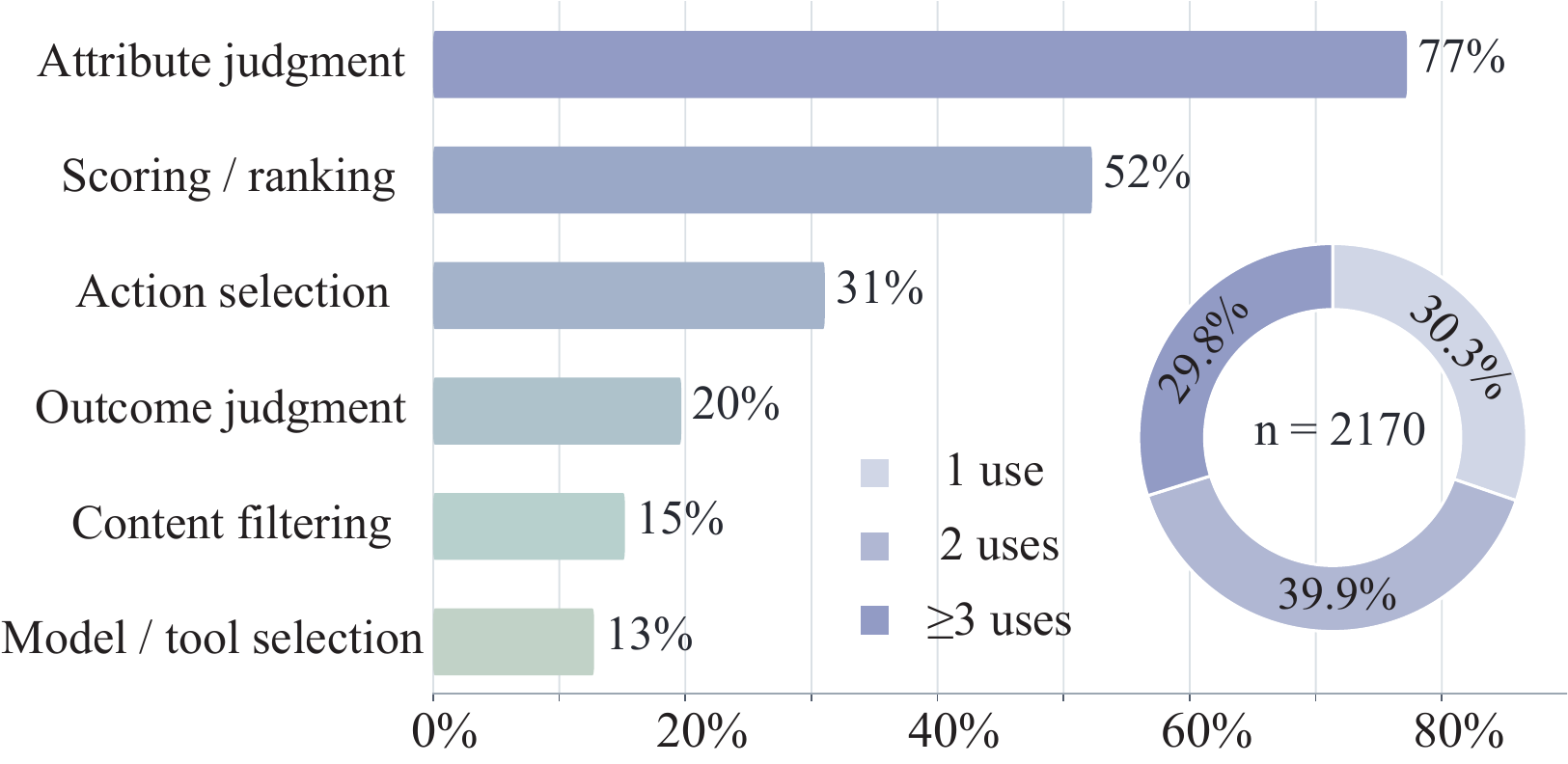}
  \vspace{-18pt}
  \caption{The bars show the share of projects for each decision purpose, and the ring shows how many purposes are identified per project. Attribute judgment is the most common purpose, while most projects with an identified purpose use Jev for more than one.}
  \label{fig:purposes}
\end{figure}

\section{Jev Usage Patterns}
\label{sec:usage}

We characterize Jev's use across domains, decision purposes, and interfaces, then examine how project activity relates to public attention.

\subsection{Application Domains}
\label{sec:domains}
Table~\ref{tab:domains} shows that Jev is used across a broad range of application domains. The two largest categories, Content \& Expert Tasks and Search \& Memory, account for only 17.8\% and 17.6\% of projects, respectively, showing that no single application dominates the ecosystem. In contrast, public attention is more concentrated. Routing \& Automation receives 41.4\% of all stars, with Model \& Tool Routing alone accounting for 32.6\%. Overall, Jev is being explored across diverse decision tasks, while community attention is particularly concentrated on routing and system integration.

\subsection{Decision Purposes and Interface Adoption}
\label{sec:purposes}

We examine what Jev decides and which interfaces projects use. A project can have several purposes and interfaces, so individual shares may overlap.

\paragraph{Decision purposes.}
Figure~\ref{fig:purposes} summarizes six decision purposes. Each bar reports the share of projects associated with a given purpose, with each project counted once per purpose, while the ring indicates the number of purposes identified in each project. Attribute judgment is the most common, appearing in 77\% of projects, followed by scoring or ranking in 52\% and action selection in 31\%. Among projects with an identified purpose, 69.7\% use Jev for two or more purposes, indicating that Jev often supports multiple types of decisions within a single application.

\begin{figure}[t]
  \centering
  \includegraphics[width=\columnwidth]{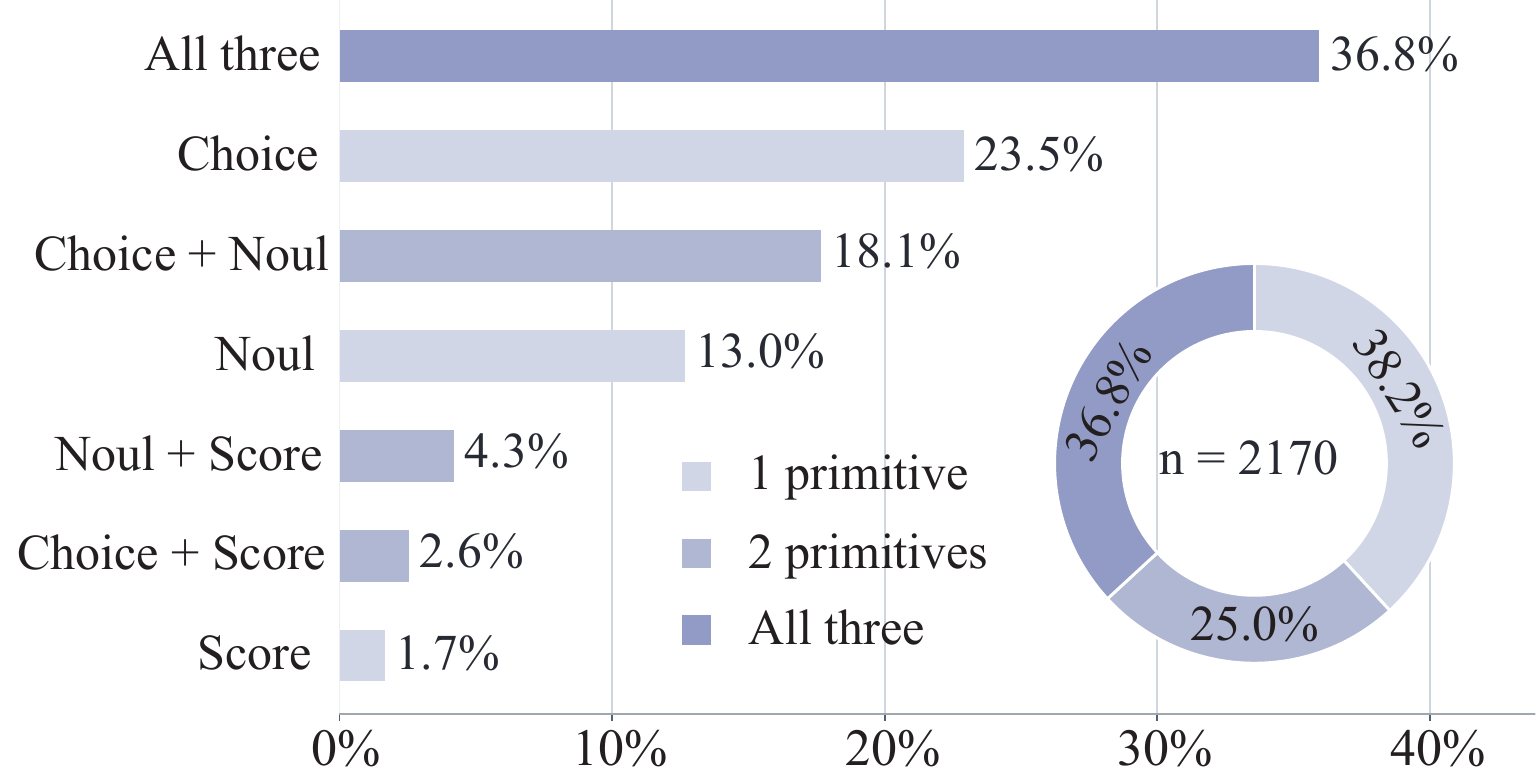}
  \vspace{-18pt}
  \caption{The bars show the share of projects using each combination of \texttt{Choice}, \texttt{Noul}, and \texttt{Score}. The ring shows how many of these interfaces are identified per project. \textbf{All three interfaces are widely used}.}
  \label{fig:interfaces}
\end{figure}

\paragraph{Interface adoption.}
Figure~\ref{fig:interfaces} compares how projects combine Jev's three interfaces. The bars give the share of projects using each exact combination of \texttt{Choice}, \texttt{Noul}, and \texttt{Score}. The ring shows whether a project uses one, two, or all three interfaces. \texttt{Choice} appears in 81.0\% of projects, \texttt{Noul} in 72.2\%, and \texttt{Score} in 45.4\%. Most projects use at least two interfaces, and 36.8\% use all three, showing that all three interfaces are widely used. Similar decision models should therefore support categorical selection, binary judgment, and scoring to facilitate developer use.

\subsection{Purpose Composition Across Domains}
\label{sec:composition}

\begin{figure}[t]
  \centering
  \includegraphics[width=\linewidth]{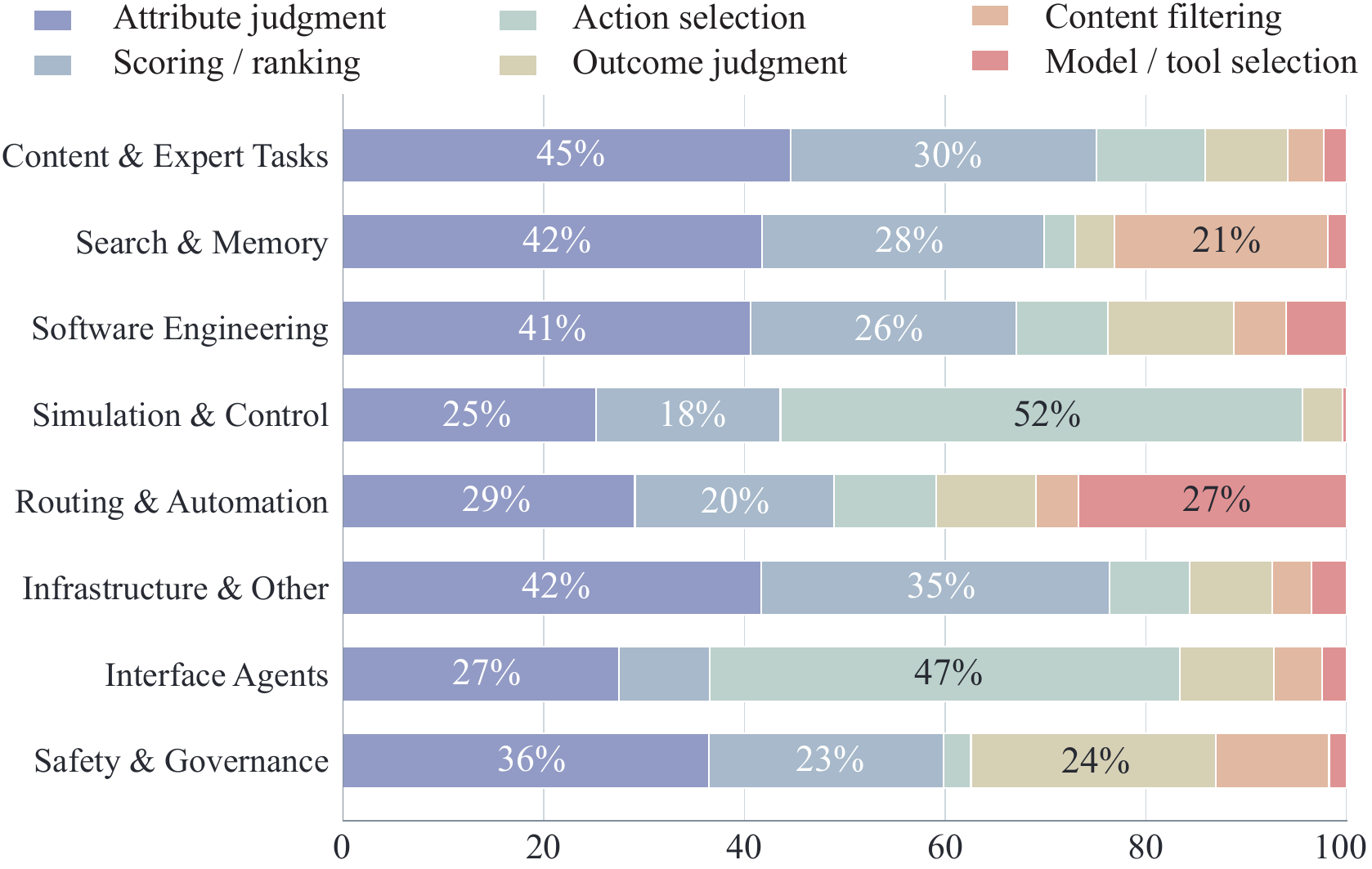}
  \caption{Decision purposes across project categories. Each bar shows the share of observed purpose labels within a category. The changing mix shows how \textbf{the surrounding workflow shapes Jev's functionality}, from judging inputs to choosing actions and tools.}
  \label{fig:composition}
\end{figure}

Figure~\ref{fig:composition} shows that Jev's role varies across project categories. Action selection accounts for 52\% of purpose labels in Simulation \& Control and 47\% in Interface Agents, reflecting Jev's use in guiding moves and interactions. Model and tool selection accounts for 27\% in Routing \& Automation but only 6\% in Software Engineering. Content filtering constitutes 21\% of labels in Search \& Memory, whereas outcome judgment accounts for 24\% in Safety \& Governance. Despite these differences, attribute judgment remains the leading purpose in six of the eight categories. Together, attribute judgment and scoring account for about 70\% of labels in Search \& Memory and 75\% in Content \& Expert Tasks. Overall, Jev serves as a reusable judgment component whose role adapts to application inputs and downstream actions.

\subsection{Project Supply and Public Attention}
\label{sec:attention}
\begin{figure}[t]
  \centering
  \includegraphics[width=\linewidth]{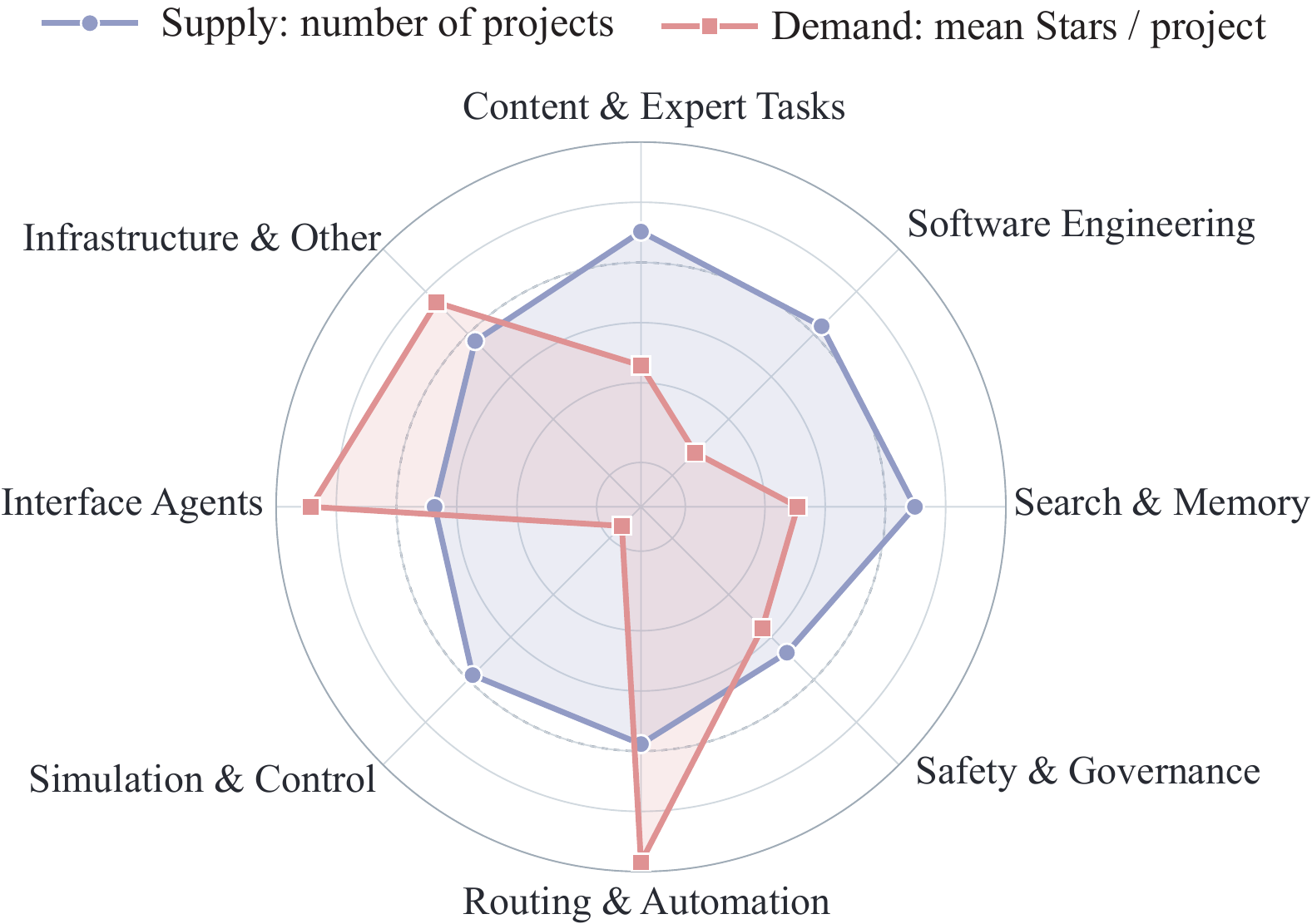}
  \caption{Project supply and public attention across eight categories. Supply is the number of projects, and demand is proxied by mean GitHub Stars per project. The two series are divided by their respective reference values of 271 projects per category and 101 Stars per project, then plotted on a logarithmic radial scale. Routing \& Automation and Interface Agents receive disproportionate attention relative to their project counts.}
  \label{fig:supply_demand}
\end{figure}

Figure~\ref{fig:supply_demand} contrasts project supply with public attention across application categories. We use project count as supply and mean GitHub stars per project as a proxy for demand. Because the measures are normalized independently, their distance from the center indicates how each category compares with the corresponding reference value.
Public attention is far more concentrated than project activity. Routing \& Automation contains 250 projects and averages 364 stars per project, while Interface Agents contains 175 and averages 271. Together, they represent only 19.6\% of projects but receive 63.0\% of all stars. By contrast, Content \& Expert Tasks and Search \& Memory are the largest categories, with 387 and 381 projects, but average just 31 and 37 stars per project. The disparity is especially clear between Routing \& Automation and Simulation \& Control. Despite nearly identical project counts of 250 and 252, they average 364 and 8 stars per project. Thus, public attention does not track project volume. Because stars measure interest in entire repositories and may be concentrated in a few prominent projects, these results indicate uneven visibility rather than unmet demand.

\section{Conclusion}
\label{sec:conclusion}
In this paper, we present a large-scale data-driven analysis of Jev's early public ecosystem, revealing rapid growth, diverse applications, and a mismatch between project distribution and public attention. Across domains, projects commonly combine multiple decision purposes and interfaces, while surrounding applications determine how model outputs guide subsequent actions. These findings suggest that general-purpose decision models should support flexible combinations of selection, judgment, and scoring. Their evaluation should cover both recurring decision purposes and the application contexts in which they are used. The concentration of public attention further motivates including less visible applications when constructing representative evaluations. Our dataset and taxonomy provide a basis for selecting these settings and studying how decision models are integrated into larger systems. Overall, this work provides a foundation for future decision-model research.

\clearpage

\section*{Limitations}

First, our analysis provides a snapshot of the Jev ecosystem as of September 22, 2026. As the ecosystem evolves, project counts, usage patterns, and public attention may shift, which could change the conclusions drawn here. Second, our dataset includes only public GitHub repositories. Private repositories and commercial applications are outside the scope of our analysis and may exhibit different usage patterns. Our findings therefore do not provide a complete account of Jev adoption.

\bibliography{custom}

\clearpage
\appendix

\section{Related Work}
\label{app:related_work}

We review the development of classification methods, existing research on Jev applications and evaluation, and closely related decision models.

\subsection{Supervised and Zero-Shot Classification}

Supervised text classifiers typically fine-tune pretrained encoders on task-specific labeled data to predict predefined categories \cite{devlin-etal-2019-bert,liu2019roberta,he2020deberta}. Zero-shot methods reduce this dependence by using natural-language label descriptions to transfer across categories and tasks \cite{yin2019benchmarking,halder2020task,clarke2023label}. Generative language models extend this flexibility through task descriptions and examples supplied in prompts, without parameter updates for each target task \cite{NEURIPS2020_1457c0d6,puri2019zero}. Instruction tuning improves generalization by training on diverse tasks expressed in natural language \cite{wei2022finetuned,sanh2021multitask}, with later work expanding task coverage and model scale \cite{wang2022super,chung2024scaling}. In practice, supervised classifiers often require additional labeled data and retraining as tasks or label sets change \cite{halder2020task}. LLMs avoid task-specific training, but autoregressive generation can incur substantial latency and inference cost \cite{NEURIPS2020_1457c0d6}. Jev targets this gap by combining zero-shot generalization with fast, low-cost inference for frequent and diverse decisions \cite{almeida2026jev,ibrahim2026evaluatingdecisionmodelstext}. Our work analyzes public Jev projects to examine how the model is used across applications and how its ecosystem is developing.
\subsection{Jev Applications and Evaluation}

Existing studies examine Jev's use within applications and its performance on decision tasks. Service orchestration studies use Jev to translate natural-language requests into structured requirements for scheduling \cite{li2026replacinglargelanguagemodels,li2026fastintentdrivenserviceorchestration}. Agent systems apply Jev to memory organization and retrieval control in Jev-Mem \cite{jiang2026jevmemsystemonecontrolledagenticmemory}, and to action selection with confidence-based fallback to a stronger LLM in REFLEX \cite{wu2026reflexjevefficientselective}. Other applications include annotating crash narratives \cite{rafe2026calibrateddecisionsscaleconverting}, selecting semantic relations before scientific calculations \cite{deng2026jevscientificdecisionsevaluating}, and estimating video quality from metadata and extracted features \cite{robitza2026jevqavideoquality}. Evaluations of text annotation and model judging assess accuracy, confidence calibration, and inference cost \cite{ibrahim2026evaluatingdecisionmodelstext,li2026jevasajudgeacceptconfidentescalate}. Their results show that accuracy and calibration vary across tasks, while routing uncertain cases to stronger LLMs can improve the balance between accuracy and cost. Robustness tests also reveal sensitivity to option names, showing that valid output types alone do not ensure correct decisions \cite{sun2026typesafeerrorfreeconstraineddecision}. Our work complements these evaluations of particular tasks and systems with a systematic data-driven analysis of Jev's public ecosystem.

\subsection{General-Purpose Decision Models}

Related models combine flexible task definitions with efficient, structured predictions. Universal classifiers use natural language inference to score label descriptions \cite{laurer2023building}, while UniMC and the Universal Discriminator select or score candidate answers across tasks \cite{wang2023zero,xu2023universal}. Compact encoders support a similar use of label descriptions: GLiNER identifies entities of user-specified types \cite{zaratiana2024gliner}, and GLiClass jointly encodes text and candidate labels for classification \cite{stepanov2025gliclass}. GLiNER2 combines classification and extraction through a shared schema \cite{zaratiana2025gliner2}, while UniEval expresses text evaluation criteria as binary questions \cite{zhong2022towards}. To reduce repeated computation, \texttt{this-that-model-1.0} returns probabilities for multiple questions in one forward pass \cite{cheng2026thisthatmodel10typeddecisionmodel}, while the Jev-inspired Visual Jev shares visual context across independent questions \cite{yu2026visualjevaccurateefficient}. Jev shares these goals through its \texttt{Choice}, \texttt{Noul}, and \texttt{Score} interfaces for software integration \cite{almeida2026jev}. These studies focus on model design and prediction quality. Our work complements them by examining how Jev is integrated into applications across its public ecosystem.

\section{Representative Applications}
\label{app:examples}

These examples show how different applications turn Jev's typed decisions into program actions. The diagrams summarize source code paths rather than recorded executions.

%
%

\begin{figure}[t]
  \centering
  \includegraphics[width=\columnwidth]{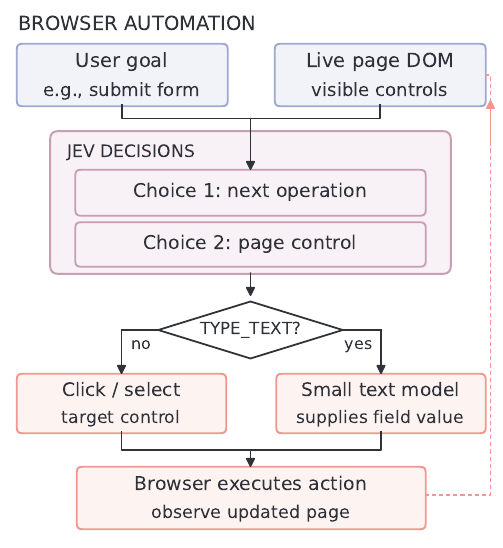}
  \vspace{-10pt}
  \caption{Jev selects a browser operation and target from the live page state. Text for typing comes from a separate model.}
  \label{fig:app_jev_ultrafast}
\end{figure}

\subsection{Jev Ultrafast: Browser Actions}

As shown in Fig.~\ref{fig:app_jev_ultrafast}, Jev Ultrafast is a browser-agent implementation that uses Jev to select actions over a structured representation of the current webpage. It illustrates how Jev can serve as a lightweight decision layer in an interactive agent loop.

\paragraph{Task and Input.}
Jev Ultrafast drives a browser toward a user-specified goal using a structured representation of the current page state \cite{jev_ultrafast_repo}. At each step, it sends Jev the goal, recent actions, page context, and an indexed list of interactive controls with their labels, current values, and supported operations. For example, a search box is represented as an editable target rather than as pixels in a screenshot.

\paragraph{Jev Decision.}
In a single request, the system submits a \texttt{Choice} question for the next operation and separate target questions for the available operation types. Jev may choose an operation such as clicking, typing, selecting an option, scrolling, waiting, or terminating the task. The agent then uses only the target answer associated with the selected operation, so that the selected target is compatible with that operation. Jev selects the action and target but does not generate the text to be entered into a field.

\paragraph{Use of the Decision.}
After validating the returned choice, the browser executor applies the selected action to the corresponding live DOM element. When the selected operation is typing, a separate small language model generates the field value from the goal and page context. This division places Jev in the role of action selector, while the browser executor performs interaction and the text model handles language generation.

\begin{figure}[t]
  \centering
  \includegraphics[width=\columnwidth]{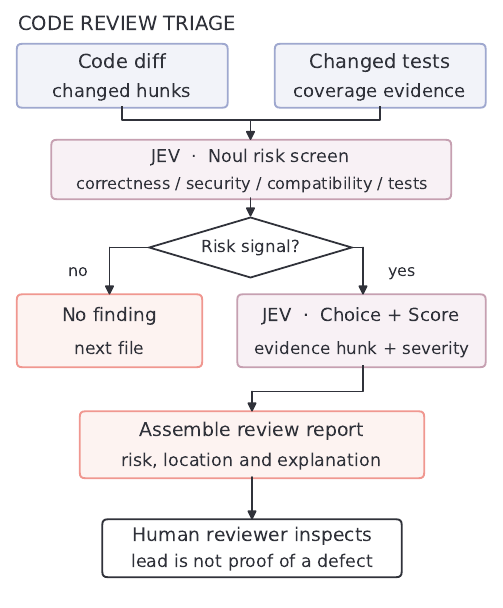}
  \vspace{-10pt}
  \caption{Risk judgments and evidence selection turn a code diff into review leads for a human.}
  \label{fig:app_jev_review}
\end{figure}

\subsection{Jev Review: Code Review Triage}

As shown in Fig.~\ref{fig:app_jev_review}, Jev Review applies Jev to code-review triage, helping identify which changes deserve closer inspection and where supporting evidence can be found. It demonstrates the use of Jev for structured risk screening and prioritization.

\paragraph{Task and Input.}
Jev Review inspects a code change to help a human reviewer focus attention \cite{jev_review_repo}. The system provides Jev with structured information extracted from the pull request, including changed files, individual patch hunks, and modified or newly added tests. These inputs allow the workflow to reason about whether a potential concern is actually supported by the submitted diff and, when relevant, identify the specific changed region that deserves closer inspection.

\paragraph{Jev Decision.}
Several \texttt{Noul} questions screen for correctness, security, reliability, compatibility, and test-coverage concerns. When a risk signal is detected, additional \texttt{Choice} questions select a supporting diff hunk and classify a possible failure mechanism. A \texttt{Score} question then estimates the priority or severity of the concern. The resulting outputs are structured judgments---for example, a risk signal, supporting code location, failure category, and severity score---rather than a free-form code review.

\paragraph{Use of the Decision.}
The surrounding program combines these structured outputs into a review report that highlights suspicious changes and directs the reviewer to the relevant evidence. Findings can therefore be ranked and inspected without requiring Jev to generate a complete natural-language review. Importantly, each reported concern remains a prompt for further inspection rather than proof of a defect: the final assessment is left to the human reviewer, together with tests, static analysis, and other validation tools. This example illustrates how Jev can serve as a triage layer that narrows the search space and prioritizes review effort within a larger software-engineering workflow.

\begin{figure}[t]
  \centering
  \includegraphics[width=\columnwidth]{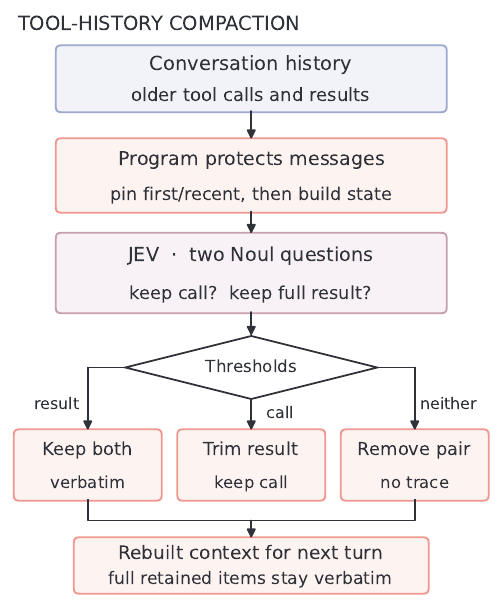}
  \vspace{-10pt}
  \caption{Two keep judgments guide whether an older tool exchange is retained, shortened, or removed.}
  \label{fig:app_fast_jev_compaction}
\end{figure}

\subsection{Fast Jev Compaction: Tool History}

As shown in Fig.~\ref{fig:app_fast_jev_compaction}, Fast Jev Compaction uses Jev to reduce the tool history retained in long-running agent conversations. Rather than summarizing previous interactions, it decides which tool calls and results remain useful for future turns.

\paragraph{Task and Input.}
Fast Jev Compaction reduces the amount of tool history carried into later agent turns \cite{fast_jev_compaction_repo}. Instead of rewriting the conversation, it presents Jev with a compact representation of the current context and evaluates eligible older tool interactions individually. Each interaction consists of a tool call and its corresponding result. Recent messages and explicitly pinned content are protected from removal, while ordinary user and assistant messages are not subject to this Jev-based filtering step.

\paragraph{Jev Decision.}
For each eligible tool interaction, two \texttt{Noul} questions ask whether remembering the tool call remains useful for continuing the task and whether the full tool result is still needed. Jev returns a keep probability for each question. The surrounding program compares these probabilities against configured thresholds, turning the model outputs into retention decisions. Importantly, Jev is not asked to summarize or rewrite the historical content; it only estimates whether different parts of the interaction should remain available.

\paragraph{Use of the Decision.}
The resulting decisions lead to three possible outcomes. If both the call and its result remain useful, the complete interaction is retained. If the call is useful but the full result is not, the call is preserved while the result is shortened. If neither is considered useful, the call--result pair is removed from the carried context. Content that survives this process remains verbatim except for the explicit truncation of selected results. This example illustrates Jev acting as a selective context-retention mechanism: it determines which parts of an agent's tool history should survive without generating a replacement summary.

\begin{figure}[t]
  \centering
  \includegraphics[width=\columnwidth]{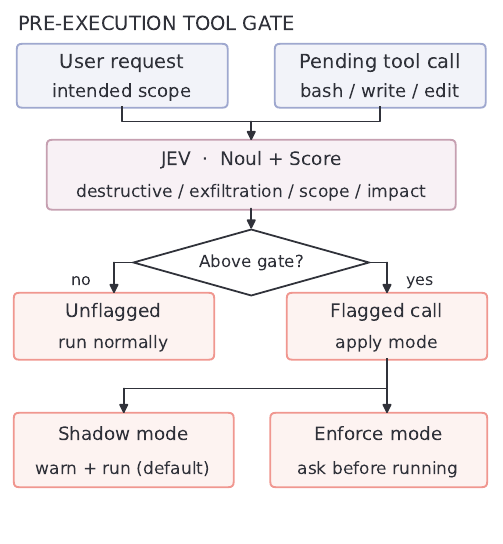}
  \vspace{-10pt}
  \caption{A typed risk assessment informs a tool gate. The extension mode controls warning or confirmation.}
  \label{fig:app_pi_jev}
\end{figure}

\subsection{pi-jev: Tool-Call Risk Gate}

As shown in Fig.~\ref{fig:app_pi_jev}, Pi-jev places Jev in front of agent tool execution as a lightweight risk gate. It evaluates whether a proposed action may be destructive, expose data, or exceed the user's requested scope before the tool is executed.

\paragraph{Task and Input.}
pi-jev checks a pending agent tool call before execution \cite{pi_jev_repo}. It sends Jev the user request, the tool name, and relevant arguments for operations such as shell commands or file edits. The intended task provides context for judging whether an operation goes beyond what the user asked.

\paragraph{Jev Decision.}
\texttt{Noul} questions assess destructive behavior, data exfiltration, and whether the operation exceeds the requested scope. A \texttt{Score} question rates potential impact. The program compares the returned probabilities and score with its configured thresholds to decide whether the call warrants attention.

\paragraph{Use of the Decision.}
In the default shadow mode, a flagged call produces a warning and can still run. In enforce mode, an interactive session asks the user to confirm before it runs. Thus Jev supplies a risk signal, while the extension's mode and policy determine whether that signal triggers a user confirmation.

\begin{figure}[t]
  \centering
  \includegraphics[width=\columnwidth]{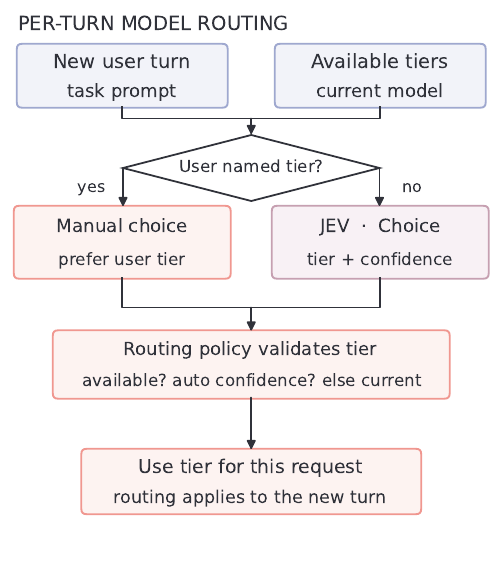}
  \vspace{-10pt}
  \caption{Jev proposes a model tier for a new turn. Routing policy applies the choice or a user override.}
  \label{fig:app_jev_router}
\end{figure}

\subsection{jev-router: Per-Turn Model Selection}

As shown in Fig.~\ref{fig:app_jev_router}, Jev-router uses Jev to select an appropriate model tier for each new user request. It demonstrates how Jev can support dynamic model routing while leaving the final routing policy to deterministic code.

\paragraph{Task and Input.}
jev-router chooses a model tier for a new assistant request \cite{jev_router_repo}. It considers the user's current task and the model tiers available to the account. The routing policy also knows the model already in use and checks whether the user explicitly named a preferred tier.

\paragraph{Jev Decision.}
For an automatically routed turn, Jev recommends a model tier together with a confidence value. Its judgment describes which available class of model fits the request. The code validates the answer against supported tiers before switching models.

\paragraph{Use of the Decision.}
The policy applies the chosen tier to the current request, subject to availability and confidence rules. An explicit user choice takes precedence, and an invalid or missing Jev answer leaves the current tier in place. This example uses Jev for task-sensitive routing while keeping final control in deterministic policy code.

\begin{figure}[t]
  \centering
  \includegraphics[width=\columnwidth]{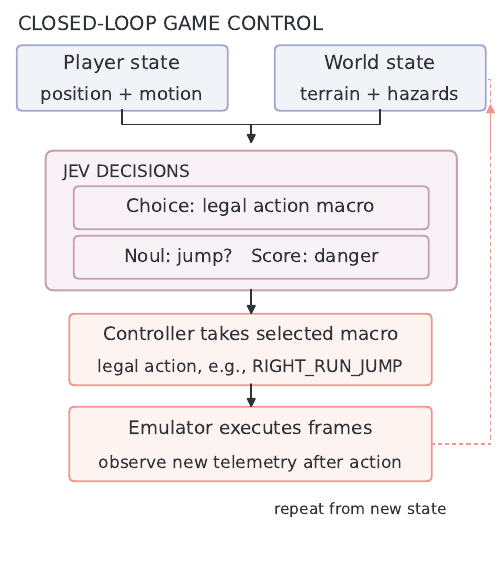}
  \vspace{-10pt}
  \caption{Structured telemetry leads to a legal action, which the emulator executes before the next observation.}
  \label{fig:app_typesafe_mario}
\end{figure}

\subsection{TypeSafe Mario: Game Control}

As shown in Fig.~\ref{fig:app_typesafe_mario}, TypeSafe Mario uses Jev as part of a closed-loop controller for a game emulator. Given structured game state, Jev selects among a small set of legal actions that are then executed in the emulator.

\paragraph{Task and Input.}
TypeSafe Mario is an experimental controller for a game emulator \cite{typesafe_mario_repo}. Instead of sending screenshots, it converts emulator telemetry and RAM into a compact structured state describing the player's position and motion, jump trajectory, nearby enemies, terrain, recent control outcomes, and timing information. Jev also receives a small predefined set of legal controller macros from which it may choose.

\paragraph{Jev Decision.}
A \texttt{Choice} question selects the next legal controller macro. A \texttt{Noul} question estimates whether a forward jump is useful at the current moment, while a \texttt{Score} question rates immediate danger. These judgments are evaluated over the same game state, but the controller action itself is determined by the \texttt{Choice}. The selected macro is therefore a discrete state-to-action decision rather than a generated game plan.

\paragraph{Use of the Decision.}
The emulator executes the selected macro for several frames, records a new structured state, and repeats the cycle. Timing calculations and state extraction remain in deterministic code, while Jev interprets the resulting state and chooses among the available actions. Jev therefore acts as a state-to-action policy inside a closed feedback loop. The project demonstrates this control integration pattern rather than providing a general benchmark of game-playing performance.

\begin{figure}[t]
  \centering
  \includegraphics[width=\columnwidth]{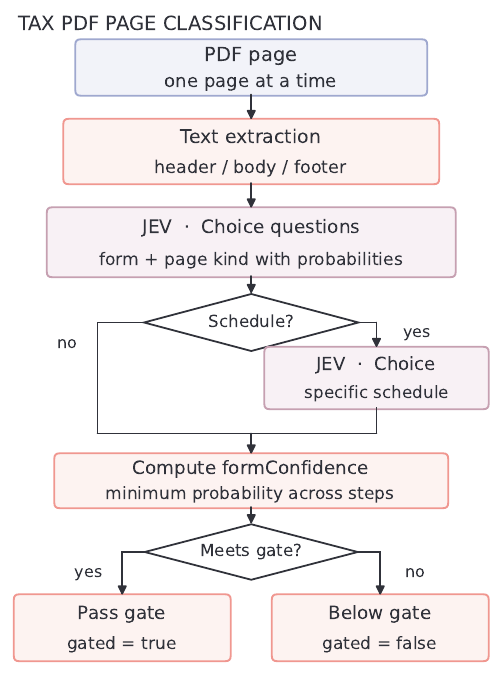}
  \vspace{-10pt}
  \caption{Page text is classified by Jev, then program code applies its own confidence gate.}
  \label{fig:app_tax_doc_classifier}
\end{figure}

\subsection{tax-doc-classifier: Tax Form Pages}

As shown in Fig.~\ref{fig:app_tax_doc_classifier}, Tax-doc-classifier uses Jev to classify individual pages of IRS tax documents. It treats Jev as a probabilistic classifier that predicts both the form identity and page type from extracted document text.

\paragraph{Task and Input.}
tax-doc-classifier identifies the IRS form and page kind of each PDF page \cite{tax_doc_classifier_repo}. A PDF text extractor provides structured page text, while a generated registry supplies descriptions of IRS forms and schedules. Each page is classified independently, and blank pages can be detected without invoking Jev.

\paragraph{Jev Decision.}
For a text-bearing page, Jev answers \texttt{Choice} questions over candidate IRS forms and seven page kinds, returning selected labels and option probabilities. Most pages are classified in one request. For several form families, a second narrower \texttt{Choice} distinguishes the parent form from its schedules. The resulting form confidence is the minimum confidence along the classification path, while the acceptance threshold is set by the surrounding program.

\paragraph{Use of the Decision.}
The program returns the predicted form, page kind, confidence, and whether the prediction passes a configurable gate. High-confidence results can be accepted for downstream processing, while lower-confidence cases can be deferred. Thus Jev provides probabilistic classification, while trust and fallback decisions remain in deterministic code.

\end{document}